\documentstyle[preprint,aps,prb]{revtex}
\begin{document}
\draft
\def\dxy{$\{311\}$}
\def\dx{$[0\bar 1 1]$}
\def\dy{$[2\bar3\bar3]$}
\def\dz{$[311]$}
\def\UI{{\it I}\ }
\def\UO{{\it O}\ }
\def\Iup{$I^{\uparrow}$}
\def\Idown{$I^{\downarrow}$}
\def\FigUC{Fig.~1}
\def\figUC{1}
\def\FigIC{Fig.~2}
\def\figIC{2}
\def\FigTI{Fig.~3}
\def\figTI{3}
\def\FigUF{Fig.~4}
\def\figUF{4}
\def\FigStep{Fig.~5}
\def\figStep{5}
\def\FigBend{Fig.~6}
\def\figBend{6}
\def\FigEfFW{Fig.~7}
\def\figEfFW{7}
\def\FigAddint{Fig.~8}
\def\figAddint{8}
\def\TabFewint{Table~I}
\def\tabFewint{I}
\def\TabEfext{Table~II}
\def\tabEfext{II}
\def\TabEfPL{Table~III} 
\def\tabEfPL{III}

\def\EqEform{1}
\def\EqEbind{2}
\def\EqEactI{3}
\def\EqPla{6}
\def\EqPlb{7}
\def\PRL{Phys. Rev. Lett.}
\def\PRB{Phys. Rev. B}
\title {\bf Extended Si defects} 
    
\author{Jeongnim Kim~\cite{JKAddress} and John W. Wilkins}
\address {Department of Physics, The Ohio State University,
Columbus OH 43210.}

\author{Furrukh S. Khan}
\address {Department of Electrical Engineering, The Ohio State University,
Columbus OH 43210.}

\author{Andrew Canning}
\address {Cray Research Swizerland, PSE, EPFL, Lausanne, Switzerland}

\maketitle
\begin{abstract}
We perform total energy calculations
based on the tight-binding Hamiltonian scheme
(i) to study the structural properties and energetics of the 
extended $\{311\}$
defects depending upon their dimensions and interstitial
concentrations and (ii) to find possible mechanisms of interstitial
capture by and release from the $\{311\}$ defects.
The generalized orbital-based linear-scaling method
implemented on Cray-T3D is used for supercell calculations
of large scale systems containing more than 1000 Si atoms.
We investigate the $\{311\}$ defects systematically from
few-interstitial clusters to planar defects.
For a given defect configuration,
constant temperature MD simulations are performed 
at 300 -- 600 K for about 1 psec 
to avoid trapping in the local minima of the atomic structures
with small energy barriers. 
We find that interstitial chain structures
along the $\langle 011\rangle$ direction
are stable interstitial defects with respect to
isolated interstitials.
The interstitial chains provide basic building blocks
of the extended $\{311\}$ defects, {\it i.e.},
the extended \dxy defects are formed by
condensation of the interstitial chains side by side
in the $\langle 233\rangle$ direction.
We find successive rotations of pairs of atoms in the $\{011\}$ plane
are mechanisms with a relatively small energy barrier
for propagation of interstitial chains.
These mechanisms together with the interstitial chain structure
can explain the growth of the $\{311\}$ defects
and related structures such as V-shape bend structures and atomic steps
observed in transmission electron microscopy images.
\end{abstract}
\pacs{61.72.Ji, 61.72.Nn, 71.15.Nc, 71.15.Fv}
\newpage
\narrowtext
\section{Introduction}

The decreasing size of semiconductor devices requires precise control
of device structures, particularly dopant distributions.   
Ion implantation introduces energetic charged
atomic  particles in a substrate for the purpose of changing
electrical, metallurgical and chemical properties  of the substrate.
The wide use of ion implantation is due to
its precise control over total
dopant doses, depth profiles and  the area uniformity.
However, ion implantation induces transient
enhanced  diffusion (TED) of dopants:
the diffusivity of the dopants is abnormally enhanced for a transient
period of time after ion-implantation.~\cite{Fa89}
Many studies have suggested that
lattice damages, introduced during ion implantation,
are responsible for fast diffusion of dopants.~\cite{Mi87,St95a}

In particular,
the diffusivity of boron during annealing in ion-implanted samples
is enhanced by many orders of magnitude greater
than B diffusivity in a thermal equilibrium.~\cite{Mi87}
This transient enhanced diffusion of B places limitations 
of the use of B$^+$-implantation in fabricating submicron devices.
Boron TED occurs due to excessive Si
interstitials, which are created by B$^+$-implantation
and contribute to the fast diffusion of B.~\cite{Fa89,St95a}
The boron diffusion via pairing of B and Si interstitials has been
supported by experiments and theories.~\cite{Fa89}
Experiments observe enhancement of the  B diffusivity 
when Si interstitials are selectively injected by surface 
oxidation~\cite{OED} and by Si$^+$-implantation.~\cite{St95a}
First principle calculations reported that the activation energy
of B diffusion associated with Si interstitials is the lowest
among possible diffusion mechanisms.~\cite{Car84a,Ni89}
Therefore, the concentration of Si interstitials 
after the ion implantation and during the thermal annealing process
is an important parameter to determine
the final depth profile of B in the ion-implanted samples.

The concentration of Si interstitials depends on (1) the ion
implantation conditions -- the incident energy and the dose -- and 
(2) the temperature at which thermal annealing is performed.
Implantation at a higher incident energy or 
with a higher dose of the ions generates more Si interstitials.
As the name transient enhanced diffusion indicates,
the enhancement factor of the B diffusivity decays as
the thermal annealing proceeds, and 
eventually the diffusivity of the
B converges to that in a thermal equilibrium.
The temporal extent of the enhanced diffusion of B decrease
at higher annealing temperatures, 
since the excessive Si interstitials
migrate to the regular lattice sites in a shorter period of time.~\cite{Mi87}
It is observed that the temporal extent (duration of the boron TED)
is exponentially activated.
The activation energy, depending on the ion implantation conditions,
shows a large variance: 1--5 eV.~\cite{Mi89,Jones1}

This activation energy summarizes complex processes for B diffusion in
ion-implanted samples -- generation, diffusion and annihilation of the
Si interstitials as well as the interactions of B with the B and Si
interstitials.
Recently, Eaglesham {\it et al.}~\cite{Ea94}
and Stolk {\it et al.}~\cite{St94}
suggested that emission of Si interstitials from 
particular extended defects, namely $\{311\}$ defects,
causes the boron TED.
They observed that the dissolution
of the $\{311\}$ defects occurs at the same time and the
temperature conditions as boron TED. 
The extended $\{311\}$ defects are detected
under B$^+$-implantation at an incident energy of few tens of
keV~\cite{St95a,St94} and the corresponding activation energy
of the boron TED was obtained at 3--5 eV.~\cite{Mi89}
The high value of the activation energy can be related
to the stability of the $\{311\}$ defects 
with respect to isolated interstitials.
However, boron TED has been also observed without detecting
any macroscopic defect, when the energy
of the  B$^+$-implantation decreases below 10 keV.~\cite{Jones1}
A lower activation energy (about 1 eV) was estimated
for this low-energy ion implantation.
This low-energy B$^+$-implantation experiment suggests
that the formation of stable $\{311\}$ defects depends on
the interstitial concentrations.
Furthermore, the observation of boron TED in the absence of
the observable $\{311\}$ defects suggests that 
microscopic interstitial clusters may exist and contribute
to the B diffusion by releasing interstitials at a lower energy cost.
It is possible that the microscopic interstitial clusters
are related to the $\{311\}$ defects and 
the formation and dissolution of the defects, from the interstitial
clusters to the extended $\{311\}$ defects,
can be explained by common mechanisms.

The proposition that the $\{311\}$ defects are formed by
condensation of interstitials and provide
interstitial sources during boron TED has been supported by
experiments using a variety of procedures to inject Si interstitials into bulk
systems: the $\{311\}$ defects are observed to be formed by
surface oxidation,~\cite{OED} and by GeV-electron
irradiation~\cite{Sali79}  and by ion implantations.~\cite{St94}

The $\{311\}$ defects are often called {\it rod-like},
because they are typically elongated 
along the $\langle011\rangle$ direction as much
as a micron.~\cite{Sali79,Fer76}
The width of the $\{311\}$ defects ranges 1 -- 100 nm along the
$\langle 233\rangle$ direction, perpendicular to the elongation
direction.
The name $\{311\}$ defects indicates the observed habit plane~\cite{Ma73}
on which the rod-like defects lie, namely the $\{311\}$ plane
formed by the $\langle 011\rangle$ and $\langle233\rangle$ directions.

Here, we present total energy calculations
based on the tight-binding Hamiltonian scheme
(i) to study the structural properties and energetics of the $\{311\}$
defects as function of their dimensions and interstitial
concentrations and (ii) to find possible mechanisms of interstitial
capture by and release from the $\{311\}$ defects.
We investigate the $\{311\}$ defects systematically from
few-interstitial clusters to planar defects.
Our results can be summarized as follows.

(1) {\it Interstitial chain -- the basic building block of the
$\{311\}$ defects:} we show that an interstitial chain along the
$\langle 011\rangle$ direction is stable with respect to isolated
interstitials. The formation energy of the interstitial chains, $
E^f_{int}=$ 2 eV per interstitial, is smaller than the formation
energies of isolated interstitials, 3 -- 5 eV.
Interstitial chains constitute basic building blocks of
the $\{311\}$ defects, {\it i.e.}, extended defects
on the $\{311\}$ habit plane can be constructed
by arranging the interstitial chains along the $\langle 233\rangle$
direction (Sec. III and IV.A).

(2) {\it Stability of the $\{311\}$ defects:}
interstitial chains along the $\langle011\rangle$ direction
are stable against the isolated interstitials
if they contain more than 2 interstitials.
The formation energy per interstitial decreases linearly with 
the length of the interstitial chains along the 
$\langle 011\rangle$ directions.
More stable extended $\{311\}$ defects than isolated
interstitial chains are formed by condensation of interstitial chains
along the $\langle 233\rangle$ direction.
The interstitial concentration of the most stable $\{311\}$ defect,
a planar defect, is $5\times 10^{14}/{\rm cm}^2$ (Sec. III, IV.B and VII).

(3) {\it Growth mechanism:}
the stability dependence of the $\{311\}$ defects 
on the interstitial concentration indicates that 
finite-size interstitial clusters capture interstitials
to grow into the interstitials chains along the $\langle 011\rangle$
direction.
The elongated interstitial chains are then further stabilized by 
capturing interstitials or interstitial chains side by side
along the $\langle 233\rangle$ direction.
This growth mechanism based on the stability study
is in a good agreement with the experimental claims
that the elongation of the rod-like $\{311\}$ defects along the
$\langle 011\rangle$ direction precedes the growth in the width
along the $\langle 233\rangle$ direction (Sec. IV.B).

(4) {\it Propagation of the interstitial chains:}
we propose a mechanism which can account for the motion of
interstitial chains in the direction perpendicular to the
chain direction with a relatively small energy barrier.
Successive rotations of pair atoms on the $\{011\}$ plane displace
the interstitial chains.
The growth of the $\{311\}$ defects along the $\langle 233\rangle$
direction can be explained by propagations of interstitial chains
which are attracted to and captured by  the 
already existing $\{311\}$ defects.
We show that V-shaped bend structure and atomic steps found
in the transmission electron microscopy (TED) images
can be formed by combinations of the interstitial chains
and the planar rotations (Sec. V and VI).
\section{Calculational details}

We perform total energy calculations based on a linear-scaling
method in a tight-binding representation,~\cite{Nscale}
using supercell methods at the $\Gamma$ point.
A tight-binding Hamiltonian developed by Kwon {\it et al.}~\cite{Kwon} 
is used to study the defect structures. 
This TB Hamiltonian 
gives a good description of the relative energies and equilibrium volumes of
the diamond structure and the metallic phases.~\cite{Flaw}
More importantly, this TB Hamiltonian 
describes the elastic properties of the diamond structure
with an error less than 5\% compared to experiments 
and gives the formation energies of the point defects
such as vacancies and interstitials
which are in good agreements with those
by the LDA calculations.~\cite{Car84b,BYJ,ChadiDB,BSC93,Thesis}
The validity and accuracy of this TB Hamiltonian
to describe the diamond structure has been addressed by several calculations
which show good agreements with the LDA calculations:
reconstructions of the Si(100) surface~\cite{JK96};
the 90$^\circ$ partial dislocations in Si~\cite{BNV96}
and hydrogenated amorphous Si.~\cite{HaSi}

The total energy calculations of defects
requires large supercells to obtain converged formation energies,
mainly due to the long-range structural relaxation.
We use the orbital-based linear-scaling method implemented
on Cray-T3D.~\cite{CGP}
We choose a spherical localization of \hbox{6 \AA\ } for
less than 1\% error in the total energy (about 30 meV per atom) 
compared to those obtained by exact calculations by diagonalizations.
The chemical potential, $\mu$, is adjusted to achieve
the correct number of electrons with an error less than 
$10^{-5}$ electron charge.
Significant charge transfer is prevented by using a finite
Hubbard-like term ($U=4$ eV); however, little differences
are observed from a finite U and $U=0$ 
in the relaxed atomic structures 
and in the total energies.

Figure \figUC\ shows 
the smallest orthorombic unit cell used to study \dxy defects. 
The $[311]$ direction, normal to the habit plane of the defects, 
is chosen as the ${\bf z}$ axis and 
the $[0\bar1 1]$ and $[2\bar3\bar3]$ directions
as the ${\bf x}$ and ${\bf y}$ axes, respectively. 
The unit lengths along the three principal axies
are $L_{xo}=a/\sqrt{2}$, $L_{yo}=a\sqrt{11}/\sqrt{2}$ and
$L_{zo}=a\sqrt{11}$, 
for the lattice constant $a$ of the diamond structure Si.
The lengths of a computational cell along the \dx and \dy directions
are varied by choosing integer multiples ($n_x, n_y$)
of $L_{xo}$ and $L_{yo}$ and
the length along the  $[311]$ direction at 
$L_z=2 L_{zo}$.
Values for $n_x$ and $n_y$ are chosen so that
the displacement of the atoms far from the defect core 
is less than 0.02 \AA\ with respect to the
regular lattice sites of the perfect diamond structure.
Periodic boundary conditions are applied along all three directions.
For the structural optimizations, initial configurations of
the model structures are given properly
and constant temperature MD simulations are performed 
at 300 -- 600 K for about 1 psec 
to avoid trapping in the local minima of the atomic structures
with small energy barriers. 
Then, atomic positions are fully relaxed by using
the steepest descent method
until the atomic force on each atom is less than 0.01 eV/\AA.
The effective temperature of the relaxed ionic configuration
is less than 0.1 K.

We define formation energy of a defect structure which contains 
$N_{int}$ interstitials and $N_{bulk}$ Si atoms
in a computational cell as
\begin{equation}
E^f = E_{tot}[N_{int}+N_{bulk}] -
{N_{int}+N_{bulk}\over N_{bulk}}E_{tot}[N_{bulk}],
\label {EFORM}
\end{equation}
The formation energy per interstitial, 
$E^f_{int}=E^f/N_{int}$,
is used to compare defects at a wide range of interstitial concentrations.
A smaller $E^f_{int}$ corresponds to a 
more stable interstitial defect.
Calculations using LDA~\cite{Car84b,BYJ,ChadiDB,BSC93} 
and using the TB Hamiltonian~\cite{Kwon}
found that the $\langle 110\rangle$-interstitialcy is the most stable point
defect with a formation energy of $3.2 - 3.9$ eV.
Our total energy calculation using a 1000-Si supercell
gives $E^f_{\langle 110\rangle}$ = 3.9 eV.

A binding energy of interstitials with respect to isolated interstitials
is defined as
\begin{equation}
-E^b = E^f - N_{int} E^f_{\langle 110\rangle}, 
\label {EBIND}
\end{equation}
for $E^f_{\langle 110\rangle}$, 
the formation energy of an isolated $\langle 110\rangle$-interstitialcy.
A positive binding energy of a defect containing interstitials 
indicates that the interstitial defect is stable
with respect to isolated interstitials.

\section{Interstitial chains along the $\langle 011\rangle$ direction}

Here, we discuss an interstitial chain structure
which is stable against isolated interstitials
and which constitutes a building block of the \dxy defects.
The interstitial chain structure in \FigIC\ shows
that 
interstitial chains along the $\langle 011\rangle$ direction 
can be inserted into bulk Si without introducing any dangling bond
by stacking pair interstitials with a periodicity of $L_{xo}$.
This arrangement of interstitials along the $\langle 011\rangle$ direction
is favorable, since only two dangling bonds would be required
for any finite-length interstitial chain.~\cite{OED,Tan1}
Since the number of dangling bonds per interstitial
is inversely proportional to the length of the chain,
the interstitial chain becomes more stable
as it grows along the $\langle 011\rangle$ direction.
The interstitial chain structures have also been used for
an atomic model of planar $\{311\}$ defects by 
Takeda {\it et al.}~\cite{TaAM,TaRev,TaSW,TaTB}

The most simple defect configuration containing interstitial chains,
the structure in \FigIC(a), is obtained by
inserting an interstitial chain into bulk Si.
By adding two interstitials per plane,
two bonds (dotted line) are broken and two seven-members
are introduced along the $\pm[311]$ directions.
The six-member ring at the center turns into two adjacent five-member rings.
The bond-angle distortion ranges from -23$^\circ$ to 20$^\circ$
and the bond-length distortion from -0.13 to \hbox{0.03 \AA}.
This chain configuration is a stable interstitial complex
with a relatively small formation energy per interstitial, 
$E^f_{int}$ = 2.2 eV,
compared to that of isolated interstitials (3-5 eV).

More stable structures can be obtained by eliminating
five-member rings which share common bonds.
A rotation of the atoms, connected by a solid bond indicated by an arrow, 
converts the right five-member ring in \FigIC(a)
into a six-member ring in \FigIC(b) and lowers the
formation energy to 1.7 eV.
This defect structure is characterized by
a six-member ring surrounded by five- and seven-member rings and
has a mirror symmetry with respect to the center of the six-member 
ring.~\cite{Note1}
While the concentration of the additional atoms is two per unit length 
($L_{xo}$),
this configuration appears to have
two interstitial chains due to its symmetry.

A rotation of the other solid bond
results in an interstitial chain surrounded by six- and five-member rings
in \FigIC(c). 
This configuration has an inversion symmetry with respect 
to the center of the interstitial pairs.
The formation energy is 1.7 eV per interstitial and
the corresponding bond-angle and bond-length distortions
are $-15^\circ$ to $20^\circ$ and -0.1 to 0.03 \AA.
We label the structure in \FigIC(b)
as an $I$-chain on $\{100\}$ plane and that 
in \FigIC(c) as an $I$-chain on $\{311\}$ plane,
based on the symmetry of the interstitial chains.
These stable interstitial chains 
constitute basic building blocks of the \dxy defects.
Arranging interstitial chains side by side in the $\langle 233\rangle$
direction results in the formation of even more stable extended defects
lying on the \dxy plane.

The small formation energy per interstitial ($< 2$ eV)
of the interstitial chain structures in Figures \figIC(b) and (c)
suggests that
microscopic defects containing only a few interstitials
may exist as stable structures as well.
Interstitial clusters containing 2--6 interstitials are constructed 
by inserting finite-length interstitial chains into bulk Si.
\TabFewint\ gives the 
formation energies per interstitial ($E^f_{int}$) 
of finite-size interstitial clusters. 
Generally, $E^f_{int}$ decreases as the size of interstitial 
defects increases, while the binding energy increases.
This $E^f_{int}$
dependence on the number of interstitials, {\it i.e.},
the length of the interstitial chain,
is consistent with
experimental observation that the rod-like defects are extended along
the $\langle 011\rangle$ direction 
as long as a submicron.~\cite{Sali79,Fer76,TaRev}
The elongation of the \dxy defects in the $\langle 011\rangle$ direction
is the consequence of 
the formation of the energetically favorable interstitial chain structures.
This observation agrees with the energetic argument 
based on the minimum dangling bond ratio of the $\langle 011\rangle$
chain structures.

Table I shows that microscopic defects containing
more than two interstitials are stable with respect to
isolated interstitials.
A binding energy per interstitial 
can be interpreted as an average energy required to release
interstitials from the interstitial clusters and the interstitial chains.
The energy cost to evaporate
interstitials increases as the size of the interstitial cluster grows:
a smaller interstitial cluster would be dissolved
at a lower annealing temperature or in a shorter period of time.
Recent experiments reported
boron TED with a small activation energy ($< 2$ eV)
even when no macroscopic defects are detected.~\cite{Jones1}.
The lack of extended defects in their samples was attributed to
the low-energy ion-implantation condition.
It is possible that few-interstitial clusters or interstitial chains,
which are stable with respect to isolated interstitials,
are generated during the low-energy ion implantation 
and contribute to the boron TED by providing interstitials
during the thermal annealing.

\section{Extended rod-like \dxy defects}

\subsection{Structure}

Here we show that extended \dxy defects can
be constructed by arranging the interstitial chains
along the $\langle 233\rangle$ direction.
It has been observed that the rod-like defects 
elongated along the $\langle 011\rangle$ direction 
grow thicker along  the $\langle 233\rangle$ direction,
when there is a constant supply of interstitials.~\cite{Fer76}
Possible configurations consisting of two interstitial chains are
shown in \FigTI.
Extended \dxy defects can be constructed by adding
more interstitial chains along the $[2 \bar3 \bar3]$ direction.
As seen in \FigTI,
the \dxy habit plane of the defects is made of two orthogonal 
directions, the $\langle 011\rangle$ and  $\langle 233\rangle$ directions.
In forming defects extended along the $[2 \bar3 \bar3]$ direction,
the arrangement of interstitial chains should satisfy following conditions:
when two interstitial chains are separated by $L_{yo}/2$ as in \FigTI(a),
the pair include an \Iup\ and an \Idown;
when separated by $L_{yo}$,
the pair consist of two identical interstitial chains,
two \Iup's as in \FigTI(b) or two \Idown's.
Otherwise, dangling bonds are introduced and
the corresponding distortions in bond angles become much larger
than typical bond-angle distortions of $\pm 20^\circ$ of the stable
\dxy defects.
Indeed, molecular dynamics simulations at 600 K about 1 psec
remove such configurations.

A new structure, an eight-member ring, is shown in \FigTI(b).
When two interstitial chains are separated by $L_{yo}$ in the \dy direction, 
there are three possible configurations:
an eight-member ring ($O_8$);
a seven-member ring ($O_7$);
and a six-member ring ($O_6$).
The three structures between two interstitial chains
are related to each other by the same kind of planar rotations in \FigIC.
The eight-member ring in \FigTI(b) can be converted
into an $O_7$  or an $O_6$ unit by rotations of the solid bonds
(heavy lines) on the $\{0\bar1 1\}$ plane.
An atomic model proposed by Takeda uses
the eight-member ring as a basic unit in constructing
the planar defects.~\cite{TaAM,TaSW,TaTB}
However, we consider three possibilities, 
the $O_8$, $O_7$ and $O_6$ units
between two interstitial chains separated by $L_{yo}$.
For all the structures we have studied,
the six-member rings ($O_6$) between interstitial chains
are found to be unstable
with respect to the $O_8$ or $O_7$ units.
A structure which forms a boundary between the defects and the bulk Si
is denoted as $E_7$ according to the notation, introduced by 
Takeda.~\cite{TaSW,TaTB}
Similarly to the eight-member rings,
the seven-member rings ($E_7$) can be converted into 
six-member rings ($E_6$) by planar rotations of the solid bonds in \FigTI.

\subsection{Energetics}

The formation energies of rod-like \dxy defects containing
few interstitial chains are listed in \TabEfext.
For example, the $E_7IIO_8IIE_7$ model structure
has the elements: $I$ -- interstitial chains; 
$O_8$ -- eight-member rings between the interstitial chains;
and $E_7$ -- seven-member rings at the defect boundaries.
Among possible combinations of the $O$ and $E$ units
for the given arrangement of interstitial chains,
configurations with a lower formation energy are
listed in \TabEfext.
The rod-like defects become more stable 
with increasing number of interstitial chains ($I$ units) in the \dy direction
as seen the $E^f_{int}$'s of \TabEfext:
the formation energy decreases 
(i) when interstitial chains are added
side by side in the \dy direction 
with a distance $L_{yo}/2$, \TabEfext\ (a) $\rightarrow$ (e); and 
(ii) when interstitial chains are added in the \dy direction
in the presence of an $O$ unit, 
\TabEfext\ (f) $\rightarrow$ (i).

As seen in \TabEfext,
the eight-member rings ($O_8$) are stable 
when they are separated by a distance equal to or larger than  $L_{yo}$.
The formation energy of the  $E_7IO_8IE_7$ model structure 
is lowered by 0.1 eV 
by transformation of the $O_8$ unit to a seven-member ring ($O_7$). 
On the other hand, 
the eight-member ring of $E_7IIO_8IIE_7$ model is stable
with respect to the $E_7IIO_7IIE_7$ structure.
Similarly, we observe that 
the seven-member ring at a defect boundary is stable,
if there are more than 2 $I$ units inserted
between $E_7$ and the closest $O$ or $E$ unit.
For example, the total energy increases
when the seven-member rings ($E_7$'s) are
converted into six-member rings for
the $E_7IIE_7$ and $E_7IIO_8IIE_7$ models.

We define an energy release $\Delta E_I$, 
when an interstitial chain 
is added to a rod-like defect $X$ 
containing $N_{int}$ interstitials per $L_{xo}$, 
as
\begin{equation}
\Delta E_I = (N_{int}\times E^f_{int}[X]+2\times E^f_{int}[E_7IE_7])
-(N_{int}+2)\times E^f_{int}[X+I].
\label {EACTI}
\end{equation}
The formation energy, $E^f_{int}[E_7IE_7] = 1.8$ per interstitial,
is used as the formation energy of an isolated interstitial chain,
which contains two interstitials per unit length $L_{xo}$.
The energy release $\Delta E_I$ 
for the $E_7IIO_8IE_7$ structure to capture
an interstitial chain and become the $E_7IIO_8IIE_7$ structure
is 1.6 eV per unit length ($L_{xo}$). 
On the other hand,  $\Delta E_I =$ 1.0 eV 
for the $E_7IIIE_7$ structure to become
the $E_7IIIIE_7$ structure.
The larger $\Delta E_I$ for the $E_7IIO_8IE_7$ structure 
indicates that the  $E_7IIO_8IE_7$ structure  provides 
a more efficient sink of interstitials
than the $E_7IIIE_7$ structure,
which contains the same number of interstitials per unit cell.
The $\Delta E_I$'s for the $E_7IIE_7$ structure
suggest that the interstitials can be evaporated with a smaller energy cost
from the  $E_7IIO_8IE_7$ structure
than from the $E_7IIIE_7$ structure.

Growth mechanism of the \dxy defects are suggested by both
the $E^f_{int}$ at different interstitial concentrations
and the energy release $\Delta E_I$ , when an interstitial chain
is added to an existing $\{311\}$ defect.
Previously  we showed that an interstitial chain can exist as a stable
structure against isolated interstitials with an $E^f_{int} < 2.2$ eV,
which is smaller than that of isolated interstitials.
The interstitial chain can provide a sink of interstitials
to capture interstitial chains along the \dy direction. 
Reactions which release a larger energy 
to capture an interstitial chain
are more likely to happen during the growth process.

We express the capturing of an interstitial chain
by a rod-like \dxy defect symbolically as
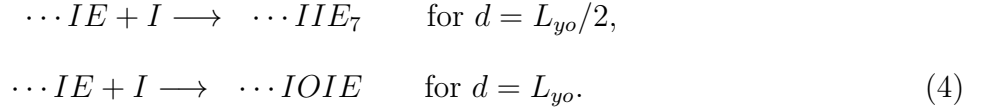
\begin{eqnarray}
\cdots IE + I \longrightarrow\ \  \cdots
IIE_7 \ \ \ \ \ \ & {\rm for}\ d = L_{yo}/2,\nonumber \\
\cdots IE + I \longrightarrow\ \  \cdots IOIE \ \ \ \ \ \ &
{\rm for}\ d = L_{yo}.\ \ \ \
\label{ICap}
\end{eqnarray}
Here, $d$ is the distance of the additional interstitial chain
from the boundary interstitial chain.
\TabEfext\ shows that the energy release 
by addition of an interstitial chain
is similar for either $d = L_{yo}/2$ or $d = L_{yo}$,
which suggests
that the interstitial-chain capturing mechanisms in Eq.~(\ref{ICap})
are equally likely to happen.
In other words,
an interstitial chain can be added next to the boundary interstitial
chain with a distance either $L_{yo}/2$ or $L_{yo}$.
Consequently, the $O$ units would be randomly introduced
and the \dxy defects would have no particular periodic arrangement
of the interstitial chains.
In fact, experiments observe no particular periodicity of the \dxy defects
in the $\langle 233\rangle$ direction,~\cite{TaRev}
while the periodicity in the $\langle 011\rangle$ direction
is identified to be $L_{xo}$, which results from 
the interstitial chain structure.

We denote the ``interstitial density'' to be
ratio of the number of interstitial {\it chains}
to the total number of units in the defects:
\begin{equation}
``{\rm interstitial\ density}"
\ \ \ \ \ \ \ \equiv\ \ \ \ \ \   {N[I]\over N[I]+N[O]+N[E]}
\label{Iratio}
\end{equation}
We obtain the optimal ratio of the interstitial chains of 67\%
for the \dxy defects based on following results:
(i) the formation energy for the extended defects (3) -- (5) 
in \TabEfext\ is same with differences less than 0.1 eV.
This formation energy, $E^f_{int} = 1.3$ eV,
is smaller than $E^f_{int} = 1.4$ eV 
of a planar defect -- an infinite sequence 
of the interstitial chains in the \dy direction 
($\cdots IIII\cdots \equiv  /I/$).
(ii) The planar defect consisting 
of sequencies of $IIO_8$ structure,
$\cdots IIO_8IIO_8\cdots \equiv /IIO_8/$ model, 
is found to be very stable with 
the smallest $E^f_{int}=$ 1.2 eV 
among the model structures we have investigated.
This interstitial density (67\%) 
is in a good agreement with the observed ratio of $I$ units,
62\% from HRTEM images.~\cite{TaAM}

\section{Propagation of interstitial chains}

So far we have discussed the energetics of classes of defects,
from the interstitial clusters to the extended \dxy defects,
solely in terms of the formation energies.
It has been shown that the interstitial clusters can
bind interstitials and become elongated interstitial chains
along the $\langle 011\rangle$ direction.
The interstitial chains in turn grow to 
the extended \dxy defects by capturing interstitials along 
the $\langle 233\rangle$ direction.
We propose that successive planar rotations, which 
introduce eight-member rings between interstitial chains
separated by $L_{yo}$,
can also displace interstitial chains.
By successive rotations of atoms connected by solid bonds
in \FigUF,
an interstitial chain characterized by the \dxy habit plane
is displaced by $\Delta = 2a/\sqrt{2}$ along the arrow
(the $[0\bar1\bar1]$ direction), 
perpendicular to the $[0\bar1 1]$ chain direction.
The planar rotations connect two interstitial chain structures
with the same $E^f_{int} = 1.7$ eV --
the $I$-chain on the \dxy plane, \FigUF(a)(c)(d),
and the $I$-chain on the $\{100\}$ plane, \FigUF(b)(e).

This kind of coordinated atomic motions was first introduced by Pandey 
as a diffusion mechanism of Si atoms in a thermal equilibrium
without introducing point defects.~\cite{Pandey}
He suggested that successive rotations of nearest neighboring atoms 
in the $\{0\bar1 1\}$ plane and the $\{011\}$ plane
result in exchanges of atoms and eventually displacements of Si atoms. 
The interstitial chain configurations 
shown in \FigUF\ are related to each other
by $\phi \sim 70^{\circ}$ rotation of the pair atoms (solid circles)
in the $\{0\bar1 1\}$ plane.

For $E^f$, a formation energy of the interstitial chain,
$E_b$, an energy barrier for the propagation, 
and $S$, an entropy involved with the propagation, 
the probability for an interstitial chain to propagate
is proportional to $\exp{\{-(E^f+E_{b}-TS)/k_BT}\}$.
The propagation mechanism of the interstitial chains
via planar rotations has such desirable features:
(1) no dangling bond is introduced during the
transformation due to atomic relaxations
and thus a small energy barrier is expected;
and (2) the entropy involved with the rotation may be
high because of the large extent of atomic relaxations
and many possible paths.

An energy barrier for the transformation from
the structure of \FigUF(a) to (b) is estimated in two ways.
First, an energy barrier for the rotation is
estimated by calculating total energies of saddle points
and comparing the total energies 
with those of the interstitial chain structures.
The calculations are performed at 
a fixed computation cell of $4L_{xo}\times 3L_{yo}\times2L_{zo}$.
This computational cell contains four solid bonds into the
$\{0\bar1 1\}$ plane which are not shown in \FigUF(a).
The saddle point configurations
are initially given by simultaneously rotating the atoms 
connected by the solid bonds by $\phi = 15^\circ, 30^\circ$, and $45^\circ$
on the $\{0\bar1 1\}$ plane.
The bond length of the two atoms is chosen at
the average value of the bond lengths of two structures in \FigUF(a) and (b).  
Then, structural relaxations of the initial configurations are performed
by allowing all atoms to move except for the two rotated atoms.
The rotated atoms are fixed at the intermediate positions.
The calculated energy barrier is  1.7 eV per bond.
When the similar rotation of two atoms is performed
in the bulk Si without any interstitial,
the energy barrier is about 2.7 eV per bond.

Alternatively,
the energy barrier is estimated by rotating
one bond at a time among four solid bonds in a computational cell,
which are underneath of the solid bond in \FigUF(a)
in the $[0\bar1 1]$ direction.
For rotations of one to three bonds,
the total energy increases by 
$1.2\pm 0.1$ eV in comparison with those of the interstitial chains
in \FigUF(a) and (b).
The corresponding increase in the formation energies
is less than 0.2 eV per interstitial.
No dangling bond is introduced
by these planar rotations
and the intermediate structures have
bond-angle distortions of $\pm 22^\circ$,
slightly larger than $-15^\circ$ to $20^\circ$
bond-angle distortions of the interstitial chain
configurations in \FigUF(a) and (b).

The energy barriers estimated by two different paths
suggest that 
many possible combinations of the planar rotations,
which has an energy barrier less than 2 eV per bond,
can lead to the displacements of interstitial chains.
The growth of the \dxy defects along the $\langle 233\rangle$ direction
during thermal annealing~\cite{St94}
can be explained by the coalescence of interstitial chains
which propagate via the planar rotations.
The same mechanism may be responsible for
the unfaulting of the \dxy defects which relates
the \dxy defects with the perfect dislocation loops
at high temperatures ($> 1000$ K).\cite{Sali89}

\section{Step structures and V-shaped bends}

We now show that step structures and V-shaped bends
found in the extended \dxy defects
can be constructed by combination of
rotations of nearest neighboring atoms 
on the $\{0\bar1 1\}$ plane 
together with the interstitial chain structures.
Rotations of three solid bonds in \FigStep(a)
result in a displacement of the right interstitial chain.
The formation energy of the structures before and after
the transformation is the same at $E^f_{int} =$ 1.5 eV.
The structure in \FigStep(b) is similar to
the core model 
of the atomic steps observed in HRTEM images.~\cite{TaRev}
Addition of interstitial chains to this defect structure
along  $\pm$ {\bf y} direction (the $[2\bar3\bar3]$ direction)
leads to the formation of a defect structure containing an atomic step.
The two \dxy habit planes shown as dashed lines are
separated by a step height of $\Delta = a/\sqrt{11}$ along the \dxy direction,
which agrees with that predicted in Ref.[25].

It has been observed that the \dxy defects change habit
planes, for example, between the $\{311\}$ and $\{3\bar1\bar1\}$ planes. 
~\cite{TaRev}
The atomic configuration around the bend 
where the habit planes change
can be identified as the $I$-chain in the $\{100\}$ plane in \FigIC(b).
We show in \FigBend\ that
the interstitial chains can be added to an $I$-chain $\{100\}$ plane
to form a $V$-shaped defect.
The initial configuration, when two interstitial chains
are added, is chosen so as to avoid
the generation of adjacent five-member rings  
by bond rotations on the $\{0\bar1 1\}$ plane.
The lowest energy configuration of the bend structure is found to
have seven-member rings at the edges ($E_7$):
by transformation of the six-member rings to $E_7$'s,
the total energy is lowered 
by 3.8 eV per cell,~\cite{Note3}
corresponding to $\Delta E^f_{int}=-0.2$ eV.
The bond-angle distortion ranges between -17 $^\circ$ and 26 $^\circ$
and the bond-length distortion between -0.1 \AA\ and 0.02 \AA. 
The formation energy, 1.3 eV per interstitial, is comparable
to that of the $E_7IIIE_7$ model structure 
at the same interstitial concentration of 6 per unit length, 
$L_{xo} = a/\sqrt{2}$.
The defect can grow on both habit planes,
the \dxy and $\{3\bar1\bar1\}$ planes, 
by capturing interstitials or interstitial chains.
The stability of the $E_7$'s suggests that
the additional interstitial chains are to be separated by $L_{yo}$.
In other words,
for the center interstitial chain $\tilde I$,
the V-shaped vend structure is
likely to have an interstitial arrangement as
$\cdots IO_8I\tilde I IO_8\cdots$.

\section{Planar \dxy defects}

Extended \dxy defects as wide as 100 nm along the 
$\langle 233\rangle$ direction
are studied by approximating them as planar defects
which are periodic both in the $\langle 011\rangle$
and the $\langle 233\rangle$ directions.~\cite{Note4}
Formation energies of planar defects,
constructed by combinations of interstitial chains ($I$) and
eight-member rings ($O_8$) along the $[2\bar3\bar3]$ direction,
are listed in \TabEfPL.
As introduced by Takeda {\it et al.},~\cite{TaSW} 
the periodicity of a planar defect along the \dy direction
is specified by an arrangement 
of $I$'s and $O$'s within a unit cell,
which is denoted by $/\cdots/$.
For example, the $/IIO_8/$ model 
contains two $I$ units separated by a distance of $L_{yo}/2$ in the \dy
direction and one $O_8$ unit within a unit cell.
The periodicity of the  $/IIO_8/$ model 
along the $[2\bar3\bar3]$ direction
is $3(L_{yo}/2)$, where $L_{yo}/2$ is the average width of each unit
and the factor 3 is the total number of units within a unit cell.

The formation energy per interstitial of the planar \dxy defects
is significantly smaller than that of isolated interstitials
and comparable to that of the extended \dxy defects with finite widths
along the $[2\bar3\bar3]$ direction (\TabEfext).
The stability of the planar defects against isolate interstitials
indicates that the \dxy defects can grow indefinitely as long as
the interstitials are supplied,
{\it e.g.}, during the ion implantation process.
Among the model structures we have investigated,
the $/IIO_8/$ model is the lowest in the formation energy per
interstitial ($E^f_{int}$ = 1.2 eV).
The density of the \UI units of the $/IIO_8/$ planar defect
is 67\%, corresponding the interstitial concentration
of $5\times 10^{14}/$cm$^2 = 4/(L_{xo}\times 3(L_{yo}/2))$.

The model structure with adjacent $O_8$ units
along the \dy direction is energetically least favorable
at the same interstitial concentration in comparing
the $/IIO_8/$, $/IIIO_8IO_8/$ and $/IIIIO_8O_8/$ models.
In fact, segments including  two adjacent eight-member rings,
$\cdots O_8O_8 \cdots$, 
have not been identified
as local structures of the extended \dxy defects.~\cite{TaAM,TaRev}
When a planar rotation is applied on the common
bond shared by the neighboring eight-member rings of 
the $/IIIIO_8O_8/$ model, 
$E^f_{int}$ is lowered by 0.2 eV.
The stability of the eight-member rings of each structure
is studied by applying planar rotations of the atoms which form boundaries
between the eight-member rings and 
six-member rings along the \dy direction.
The last column of \TabEfPL\ shows that 
the $O_8$'s are stable when their separation is equal to or
larger than $L_{yo}$, consistently with
the result of the finite-width \dxy defects.

The difference of the $E^f_{int}$'s of 
the $/IIO_8/$ and $/IIIO_8IO_8/$ models
implies that the stability of the planar defects
is determined not only by the density of interstitial chains
but also the arrangement of interstitial chains.
Planar defects containing isolated interstitial chains
between $O$ units ($\cdots OIO\cdots$)
have higher formation energies than those without 
them at the same interstitial concentration.
The energy gain by transformation of one of the eight-member rings
to a seven-member ring is insufficient to change 
the relative stabilities of the planar defects.
Based on the $E^f_{int}$ of the planar defects listed in \TabEfPL, 
we expect that the $/IIIO_8IIO_8/$ and $/IIIO_8/$ planar defects
to have comparable $E^f_{int}$ to the most stable planar defect,
the $/IIO_8/$ model. 

The formation energy of the planar defects we have discussed 
is obtained without 
allowing relaxation of the supercells.
A supercell relaxation would have generated 
the displacement vector (Burger vector)
around the core of the
\dxy defects identified by TEM observations.~\cite{Fer76,TaAM}
However, we anticipate that the {\it relative stability} 
is correctly 
described by the fixed cell calculations.
The critical dimension to obtain the converged formation energy
for the planar defects
is $L_z$, the length of the supercell in the $[311]$ direction.
We obtain the formation energies of the planar defects
at $L_z = 3L_{zo}$ and the formation energies are also listed in \TabEfPL.
The formation energy is lowered by 0.08 eV on average
for all the planar defects.
The relative stability of the planar defects 
is the same for $L_z = 2L_{zo}$
and $3L_{zo}$.

In \FigEfFW, we show the formation energy of the defects
which are infinite along the $[2\bar3\bar3]$ direction
and finite ($L_D = n_D L_{xo}$) along the $[0\bar1 1]$ direction.
The elongation direction of the \dxy defects is always
along the $\langle 011\rangle$ direction.
However, we study 
the hypothetical defect structures with lengths ($L_D$)
shorter than the width in the \dy direction
to compare with the planar defects.
Since the infinite limits of these defects ($L_D \propto \infty$)
are planar defects,
we can compare the formation energy of the planar defects in \TabEfPL\  
with $E^f_{int}[\infty]$, which can be extracted 
from the finite-length-defect calculations.

The defect structures in \FigEfFW\ are 
constructed by inserting a finite-length planar defect
between bulk layers.
The defects become
either the $/I/$ or $/IO_8/$ model structure
at the infinite limits. 
The formation energy increase linearly
to the length along the $[0\bar1 1]$ direction:
\begin{equation}
E^f = E^f_{edge} + N_{int} E^f_{int}[\infty].
\label{Eflinea}
\end{equation}
Then the $E^f_{int}$ of the defects with respect to the length is written as
\begin{equation}
E^f_{int} = E^f_{int}[\infty] + E^f_{edge}/N_{int} \propto 1/L_D 
\label{Eflineb}
\end{equation}
The slopes, $E^f_{int}[\infty]$, of two kinds of defects is
the formation energy per interstitial at the infinite limits, {\it i.e.},
$N_{int} \propto L_D \rightarrow \infty$.
The slopes obtained from linear fittings
are in good agreements with the $E^f_{int}$'s in \TabEfPL,
1.4 eV and 1.7 eV for the $/I/$ model and $/IO_8/$ model, respectively.
This result confirms that the $/I/$ model is more stable
than the $/IO_8/$ model.

Finally, the ability of the \dxy defects to capture or release
interstitials is studied by calculating an energy required
to add an interstitial to the bulk Si in the presence
of the planar defects (\FigAddint).
An interstitial is added at the hexagonal site at 
a distance of $h$ from the \dxy habit plane.
The eight-member rings of the $/IO_8/$ planar defect
can provide efficient sinks of interstitials:
the extra energy to add an interstitial to the $/IO_8/$ planar defect
decreases with decreasing $h$, 3.2 eV at 16.5 \AA\ and 1.5 eV
at 10.1 \AA. 
When the interstitial is added right above the eight-member
ring (3.8 \AA\ from the habit plane),
about 0.1 eV energy is released.
On the other hand, an additional interstitial is not
bounded by the $/IIO_8/$ planar defect.
The energy required to add an interstitial is  4.7 -- 5.1 eV.
This energy is comparable to the formation energy of an isolated
interstitial at the hexagonal sites
and larger than that of the $\langle 110\rangle$ interstitialcy.
These results support the calculations
showing that the saturation of interstitials is achieved
at the $I$ density about 67\% in the $/IIO_8/$ structure (\TabEfPL).

\section{Conclusions}

We have showed that interstitial chain structures elongated
along the $\langle 011\rangle$ direction
constitute the basic building blocks
of the extended \dxy defects.
This interstitial chain configuration is favored due to
the minimum ratio of dangling bonds per interstitial
and small distortions in bond angles and bond lengths.
Even finite-size interstitial chains,
as small as 2-interstitial clusters, are found to be stable
against isolated interstitials.
A growth mechanism of the \dxy defects
can be speculated from the dependence of 
the $E^f_{int}$'s on the dimensions of the defects
and the interstitial concentrations.
The growth  of the \dxy defects would occur first by
(1) elongation of interstitial chains
along the $\langle 011\rangle$ direction 
and (2) the widening of the defects follows
by capturing interstitial chains
alongside the $\langle 233\rangle$ direction.
We present an efficient mechanism, 
rotations of atoms on the $\{011\}$ plane, 
which can lead to propagation of interstitial chains.
The planar rotation together with the interstitial chain structures
can be applied to explain growth process of the \dxy defects
and related structures such as the V-shape bend structure
and atomic steps.
We observed that the most stable arrangement of interstitial chains
of the extended \dxy defects is $\cdots IIO_8IIO_8\cdots$
and the optimal density of interstitial chains is 67\% 
which is in a good agreement with the experimental value of 62\%.

{\centerline {\bf Acknowledgements}}

We would like to thank B.~Albers, 
J.~Wills, K.~Jones, T. Lenosky and G.~Galli for useful discussions
and L. J\"onsson for critical readings of the manuscript.
This work is supported by Department of Energy --
Basic Sciences, Division of Materials Sciences.
The computational facilities are provided by the
Ohio Supercomputer Center and the Pittsburgh Supercomputer Center.
The parallel implementation of the generalized $O(N)$ method
is developed 
as part of the Parallel Application Technology Program (PATP) joint
project between \'Ecole Polytechnique F\'ed\'erale de Lausanne (EPFL)
and Cray Research.
\newpage


%
\newpage
\figure { {\bf\FigUC} 
~An orthorombic unit cell used to describe
the $\{311\}$ defects. The three principal axes are along
the $[0 \bar 1 1]$, $[2 \bar3 \bar3]$ and $[311]$ directions.
For the Si lattice constant  ($a=5.43$ \AA),
the dimensions of the unit cell are
$L_{xo} = a/\sqrt{2}, L_{yo} = a \sqrt{11}/\sqrt{2}$ and $L_{zo}=a \sqrt{1
1}$.
Supercell calculations at the $\Gamma$ point are performed
by taking a computational
cell whose size is
$L_x = n_x\times L_{xo}, L_y = n_y\times L_{yo}$,
and $L_z = 2\times L_{zo}$.
Periodic boundary conditions are applied in all three directions.
}

\figure { {\bf\FigIC}~An interstitial chain obtained
by stacking pair interstitials
with a periodicity of $L_{xo}$ along the $\langle 011\rangle$ direction.
The solid atoms are $L_{xo}/2$ into the plane with respect
to the open-circled atoms.
(a) An interstitial chain inserted into bulk Si
is surrounded by two adjacent five-member rings.
The dotted lines denote the bonds which are broken by
addition of interstitial chains.
This structure has a formation energy of 2.2 eV per interstitial.
(b) By a rotation of atoms connected by a solid bond (arrow),
a more stable structure with $E^f_{int} =$ 1.7 eV is obtained.
The structure has a mirror symmetry with respect to the center
of the six-member ring.
The dashed line indicates the $\{100\}$ plane normal
to the center line.
(c) A rotation of the other bond
results in an interstitial chain configuration of $E^f_{int}=$ 1.7 eV.
The symmetry of the structure is
an inversion symmetry with respect to the center of pair interstitials
on the $\{311\}$ plane (dashed line).
}

\figure { {\bf\FigTI}~Projections on the $\{0\bar1 1\}$ plane
of the rod-like \dxy defects containing two interstitial chains.
The solid atoms ($\bullet$)
are deeper into the plane than
the open-circled atoms ($\circ$) by $L_{xo}/2$.
Interstitial chains are specified by the upper atoms with respect
to the $\{311\}$ plane (dashed lines): (1) \Iup,
when the upper atoms are out-of the $\{0\bar1 1\}$ plane;
and (2) \Idown, when the upper atoms are into the $\{0\bar1 1\}$ plane.
The rod-like \dxy defects are believed to grow by adding interstitial
chains along the \dy direction.
Total energy calculations give the formation energy per interstitial
at $1.5$ eV for (a) and at $1.6$ eV for (b).
Seven-member rings ($E_7$'s) form boundaries between the defects
and the bulk Si, and they can be converted into six-member rings ($E_6$)
by rotations of the solid bonds (heavy lines).
In a similar way, an eight-member ring $O_8$ can be converted
into either a seven-member ring ($O_7$) or a six-member ring ($O_6$).
The stability of the $E_7$ and $O_8$ units depends on the
arrangement of interstitial chains.
A rotation of any solid bond of (a) increases the total energy.
However, more stable structures can be obtained by
rotations of the solid bonds in (b) as far as
the transformation doesn't introduce
adjacent five-member rings.
}

\figure { {\bf\FigUF}~By succesive rotations of
solid atoms connected by solid bonds,
an interstitial chain on the $\{311\}$ habit plane is displaced by
$\Delta = 2 a/\sqrt{2}$ along the perpendicular direction
of the interstitial chain (arrow).
The $\otimes$ denotes a reference atom.
Intermediate structures are characterized by
the habit planes, the $\{100\}$ plane for (b) and (d)
and the $\{311\}$ plane for (c), indicated by dashed lines.
The formation energy per interstitial of the illustrated structures
is 1.7 eV for a fixed computational cell.
The dimension of the cell is $4L_{xo} \times 3L_{yo}\times 2L_{zo}$.
}

\figure { {\bf\FigStep}~By rotations of three solid bonds,
a rod-like \dxy defect containing two $I$-chains
on the \dxy plane (a) is transformed to
a step configuration with a height of $\Delta = a/\sqrt{11}$
along the $[311]$ direction (b).
The solid atoms are deeper into the plane by $L_{xo}/2$
than the open-circled atoms.
The difference of the total energies per unit cell of the two structures,
which contains 528 bulk Si atoms,
is less than 0.1 eV, and the
corresponding $\Delta E^f_{int}$ is less than 0.05 eV.
Intermediate states after rotations of one or two solid bonds
have the same $E^f_{int}$.
}

\figure { {\bf\FigBend}~Bend configuration of the \dxy defects (bottom)
obtained by addition of two interstitial chains to
an $I$-chain on the $\{100\}$ plane (top).
Insertion of interstitial chains results in
breaking of bonds (dotted lines).
The bottom structure contains 6 interstitials per unit length of $L_{xo}$:
2 interstitials associated with the $I$-chain on the $\{100\}$ plane
and 4 additional atoms indicated by arrows.
A row of the center six-member rings along the $[0\bar1 1]$ direction
forms a boundary of
the defects which have the $\{311\}$ and $\{3\bar1\bar1\}$
habit planes (dashed lines).
This V-shaped defect has a mirror symmetry with
respect to the center of the six-member rings.
The formation energies per interstitial
are 1.7 eV (top) and 1.3 eV (bottom).
Further growth of the defect can occur by
addition of interstitial chains on the both habit planes.
}

\figure { {\bf\FigEfFW}~Formation energy ($E^f$) per unit width ($L_{yo}$)
with respect to the number of interstitials per unit width (left);
a schematic diagram to show how the computational cell is constructed (rig
ht).
A slab with a finite length of $n_DL_{xo}$, for an integer $n_D = 1 \cdots
 4$ ,
contains defect structures
based on the $/IO_8/$ and $/I/$ models.
The slab is sandwiched between Si bulk layers along the $[0\bar1 1]$
direction  with a fixed total length of $6L_{xo}$.
A common width along the $[2\bar3\bar3]$ direction is chosen at $L_{yo}$.
The number of interstitials per unit width ($L_{yo}$)
is proportional to the length of the defects:
$N_{int} = 4 n_D$ for the defects based on
the $/I/$ model
and $N_{int} = 2 n_D$ for those based on the $/IO_8/$ model.
The formation energy is fitted by Eq.~(\EqPla).
For the defects based on the $/I/$ model,
the values of $E^f_{edge}$ and
$E^f_{int}[\infty]$ are 8.4 and 1.4 eV respectively;
$E^f_{edge} = 6.1$ eV and $E^f_{int}[\infty] = 1.7$ eV
for the defects derived from the $/IO_8/$ model.
The formation energy per interstitial decreases as the length increases
as in Eq.~(\EqPlb).
At $N_{int}\rightarrow \infty$, the $E^f_{int}$'s
converge to those of the planar defects, 1.4 eV for the $/I/$ model
and 1.7 eV for the $/IO_8/$ model.
}
\figure { {\bf\FigAddint}~Initial configurations
to obtain the extra energy required to add {\it one} interstitial
to the bulk system in the presence of a planar defect.
The additional interstitial is located
at a hexagonal site with a distance
$h$ measured from
the $\{311\}$ plane on which the center of interstitial chains lie
(the dashed line).
Two cases with $h_1 = 16.5$ \AA\ and $h_2 = 3.8$ \AA\ are
illustrated at the same time.
The computational cell is $6L_{xo}\times 3L_{yo}\times 2L_{zo}$.
The distance of interstitials in the neighboring cells,
{\it i.e.} the distances between images
related by periodic boundary conditions,
is larger than the distance of the additional interstitial
to the habit plane of the planar defect.
The interstitials associated with the interstitial chains
of the $/IO_8/$ model structure are indicated by arrows.
}
%
%
\begin{table}
\begin{tabular}{c|ccc|ccc}
          & \multicolumn{3}{c|}{$I$-chain on $\{311\}$ plane $^a$}
          & \multicolumn{3}{c}{$I$-chain on $\{100\}$ plane $^b$}\\
$\ \ \ \ \ N_{int}\backslash$ eV \hfil& 
$E^f/N_{int}$  &  $E^b$ & $E^b/N_{int}$ & 
$E^f/N_{int}$  &  $E^b$ & $E^b/N_{int}$ \\
\hline
            2 & 4.7 & -1.6 & -0.8 & 4.9 & -2.0 & -1.0 \\
            3 & 3.7 &  0.6 &  0.2 & 3.4 &  1.5 &  0.5 \\
            4 & 3.4 &  2.0 &  0.5 & 3.3 &  2.4 &  0.6 \\
            5 & 3.0 &  4.5 &  0.9 & 2.9 &  5.0 &  1.0 \\
            6 & 2.8 &  6.6 &  1.1 & 2.8 &  6.6 &  1.1 \\
     $\infty$ & 1.7 &      &  2.2 & 1.7 &      &  2.2 \\
\end{tabular}
Orientation of the computational cell:
$^a [0\bar1 1]\times [3\bar2\bar2]\times [311]$

\hskip2.77truein $^b [0\bar1 1]\times [0\bar1\bar1]\times [100]$
\vskip 0.5truecm
\caption{Formation energy per interstitial,
$E^f_{int}$ in Eq. (\EqEform), and
the binding energy, $E^b$ in Eq. (\EqEbind), of few-interstitial clusters.
The interstitial clusters are constructed by inserting interstitial chains
between bulk layers whose length is at least
$6\times L_{xo}$ along the $[0 \bar1 1]$ direction.
The length of an $N_{int}$--interstitial
cluster along the $\langle 011\rangle$
direction is $N_{int} L_{xo}/2$.
The orientation of the computational cells
is chosen according to  the symmetry of the interstitial chains.
The bottom row shows $E^f_{int}$ and
$E^b/N_{int}$ for the line defects-
infinite interstitial chains along the $[1\bar1 0]$ direction (\FigIC).
The size dependence of $E^f_{int}$'s shows that
the interstitial clusters can trap interstitials
to form more stable elongated structures.
}
\label{tab1}
\end{table}

\begin{table}
\begin{tabular}{lccc}
& $N_{int}$  (\% of $I$ units) &  $E^f_{int}$ & $\Delta E_I$
:$X+I \rightarrow Y$\\
Model structure & per unit cell$^*$& eV & \\
\hline
(a) $E_7IE_7^{**}$      &  2  (67)  & 1.8$^\dagger$
                        & 1.2: (a)$+I \rightarrow$ (b)\\
                    & & & 0.9: (a)$+I \rightarrow$ (f)\\
(b) $E_7IIE_7$          &  4  (50)  & 1.5
                        & 1.6: (b)$+I \rightarrow$ (c)\\
                    & & & 1.0: (b)$+I \rightarrow$ (g)\\
(c) $E_7IIIE_7$         &  6  (60)  & 1.3
                        & 1.2: (c)$+I \rightarrow$ (d)\\
                    & & & 0.6: (c)$+I \rightarrow$ (h)\\
(d) $E_7IIIIE_7$        &  8  (67)  & 1.3
                        & 1.0: (d)$+I \rightarrow$ (e)\\
(e) $E_7IIIIIE_7$       & 10 (71)    & 1.3       & \\
(f) $E_7IO_7IE_7$ or $E_7IO_8IE_6$       &  4  (40)  & 1.6
                        & 1.3: (f)$+I \rightarrow$ (g)\\
(g) $E_7IIO_7IE_7$ or $E_7IIO_8IE_6$      &  6  (50)  & 1.4
                        & 1.1: (g)$+I \rightarrow$ (h)\\
                    & & & 1.3: (g)$+I \rightarrow$ (i)\\
(h) $E_7IIIO_7IE_7$ or $E_7IIIO_8IE_6$     &  8  (57)  & 1.3       & \\
(i) $E_7IIO_8IIE_7$     &  8  (57)  & 1.3       & \\
\end{tabular}
$^*$ Unit cell: $L_{xo}\times 6L_{yo}\times 2L_{zo}$

$^{**}$ Identical to the $I$-chain on the \dxy plane in \FigIC(c).

$^\dagger$ $E^f_{int} = 1.7$ eV for a unit cell
$L_{xo}\times {\bf 3}L_{yo}\times 2L_{zo}$
\vskip 0.5truecm
\caption{Formation energy of extended
\dxy defects consisting of interstitial chains.
The computational cell is $3L_{xo}\times 6L_{yo}\times 2L_{zo}$
containing three unit cells in the $[0\bar1 1]$ direction
for all listed model structures.
The ratio of the $I$ unit to the extent of defects
is given in percentage, e.g., 3/6 = 50 \% for the $E_7IIO_8IE_7$ structure.
The energy release, $\Delta E_I$, is defined in Eq. \EqEactI\ 
upon adding an interstitial chain ($I$) to
the rod-like defect $X$.
$Y$ denotes a defect structure constructed by
adding an $I$ with a distance of either $L_{yo}/2$ or $L_{yo}$
from the boundary interstitial chain of the defect $X$.
}
\label{tab2}
\end{table}

\begin{table}
\begin{tabular}{ccccc}
Model structure & $N_{int}^*$   & \multicolumn{2}{c}{$E^f_{int} (eV)$}
    & Stability of $O_8$ units\\
& & $2L_{zo}$ & $3L_{zo}$ &\\
\hline
       $/I/$      & 12 & 1.35 & 1.27 & \\
       $/IO_8/$     &  6  & 1.68 & 1.59 & Unstable, $/IO_7IO_7IO_8/$ \\
       $/IIO_8/$    &  8  & 1.23 & 1.16 & Stable \\
       $/IIIO_8IO_8/$ &  8  & 1.35 & 1.26 & Unstable, $/IIIO_7IO_8/$ \\
       $/IIIIIO_8/$ & 10 &  1.24 & 1.18 & Stable \\
       $/IIIIO_8O_8/$ & 8  & 1.48 & & Unstable, $/IIIIO_7O_7/$ \\
\end{tabular}
$^*$ Per unit area: $L_{xo}\times 3L_{yo}$
\vskip 0.5truecm
\caption{Formation energies  per interstitial ($E^f_{int}$)
of planar \dxy defects
consisting of interstitial chains and eight-member rings.
The periodicity of the structures along the \dy direction is denoted by
\hbox{{\it /$\cdots$/}}.
The number of interstitials and the formation energy
are given per unit area, $L_{xo}\times 3L_{yo}$,
in order to compare structures with different periodicities
along the \dy direction.
The same computational cell is used for the total energy calculations
of all the listed structures with two choices for
the length of the cell in the $[311]$ direction, $2L_{zo}$
and $3L_{zo}$.
The last column shows the stability of the eight-member rings
and the most stable planar defects
which can be obtained by rotations of atoms on the $\{0\bar1 1\}$.
}
\label{tab3}
\end{table}

\end{document}